\input{aipcheck}

\documentclass[
    ,final            
  ]
  {aipproc}

\layoutstyle{6x9}

\usepackage{amsmath}

\begin{document}

\title{Nonlinear Propagation of Crossing Electromagnetic Waves in Vacuum due to Photon-Photon Scattering}


\author{Daniele Tommasini}{
  address={Departamento de F\'\i sica Aplicada. Universidade de Vigo.
As Lagoas E-32004 Ourense, Spain.}
}

\author{Albert Ferrando}{
  address={Interdisciplinary Modeling Group, InterTech. Departament d'\`{O}ptica,
Universitat de Val\`{e}ncia. Dr. Moliner, 50. E-46100 Burjassot (Val\`{e}ncia),
Spain.}
}

\author{Humberto Michinel}{
  address={Departamento de F\'\i sica Aplicada. Universidade de Vigo.
As Lagoas E-32004 Ourense, Spain.}
}

\author{Marcos Seco}{
  address={Departamento de F\'{\i}sica de Part\'{\i}culas. Universidade de Santiago de Compostela.
E-15706 Santiago de Compostela, Spain.}
}

\begin{abstract}
We review the theory for photon-photon scattering in vacuum,
and some of the proposals for its experimental search,
including the results of our recent works on the subject.
We then describe a very simple and sensitive
proposal of an experiment and discuss how it
can be used at the present (HERCULES)
and the future (ELI) ultrahigh power laser
facilities either to find the first evidence of
photon-photon scattering in vacuum,
or to significantly improve
the current experimental limits.
\end{abstract}

\maketitle

Among other interesting phenomena \cite{weinberg,tommasini0203},
Quantum Electrodynamics (QED) and non-standard theories of the electromagnetic
field\cite{BornInfeld,Denisov} predict that the vacuum should behave like
a kind of {\it virtual} electron-positron plasma, thus allowing for
photon-photon ($\gamma\gamma$) scattering.
However, the latter prediction has not yet been confirmed by experiment,
not even indirectly. In fact, two possible strategies can be used for the quest of
$\gamma\gamma$ scattering.
On one hand, the cross section for the process will be maximum
at a possible future photon-photon
collider\cite{phcollider}, based on an electron laser
producing two beams of photons in the $MeV$ range
(i.e., having wavelengths in the range of a few $fm$).
A second approach will be to perform experiments using ultrahigh power optical lasers,
such as those that are already in use today\cite{mourou06,HERCULES} or
will be available in the near future\cite{mourou06,ELI},
in such a way that the high density of
photons will compensate the smallness of the cross section.
In this case, the photon energies are well below the electron
rest energy, and the effect of photon-photon
collisions due to the interchange of virtual
electron-positron pairs can be expressed in terms of
the effective Euler-Heisenberg nonlinear Lagrangian\cite{Halpern,Heisenberg}.
This modifies Maxwell's equations for the average values of the electromagnetic
quantum fields\cite{mckenna} and affects the properties of the QED vacuum\cite{klein}.

Here, we will briefly review the theory for $\gamma\gamma$ scattering,
and discuss some of the proposals for its experimental search.
We will then consider the present and future ultrahigh
power laser facilities, and show that, besides their
important technological uses, they will provide a unique opportunity
for the search of $\gamma\gamma$ scattering, taking into account
the results of our recent works \cite{prl99,tommasinietal08}.
Finally, we will describe a very simple and sensitive
proposal of an experiment and discuss how it
can be used at the present (HERCULES) \cite{HERCULES}
and the future (ELI) \cite{ELI} ultrahigh power laser
facilities either to find $\gamma\gamma$ scattering, or to improve significantly
the current experimental limits.

\section{The QED nonlinear wave equation}
Even in the absence of matter, photons can interact with
each other by interchanging virtual electron-positron pairs in a loop.
In this sense, the vacuum behaves like some kind of virtual
electron-positron plasma.
We will assume that the photon energy is well below the
threshold for the production of electron-positron pairs,
$2 m_e c^2\simeq 1 MeV$. This means that we will only
consider radiation of wavelengths $\lambda\gg 2\times 10^{-13} m$,
which is always the case in optical experiments.
In this case, the QED effects can be described by the
Euler-Heisenberg effective Lagrangian density\cite{Heisenberg,weinberg},

\begin{equation}
{\cal L}={\cal L}_0
+\xi_l {\cal L}_0^2+ \xi_t \epsilon_0^2 c^2({\bf E}\cdot {\bf B})^2,
\label{L}
\end{equation}
being
\begin{equation}
{\cal L}_0=\frac{\epsilon_0}{2}\left({\bf E}^2-{c^2\bf B}^2\right)
\label{L0}
\end{equation}
the linear Lagrangian density and $\epsilon_0$ and $c$ the dielectric
constant and the speed of light in vacuum, respectively. Here,
${\bf E }$ and ${\bf B}$ are the expectation values of the electromagnetic fields,
that can be measured by the usual classical means e.g. using test charges.

The additional, nonlinear terms, that appear multiplying the parameters
$\xi_l$ and $\xi_t$ in Eq. \eqref{L}, are the only two
Lorentz-covariant terms that can be formed with the electromagnetic fields
at the lowest order above ${\cal L}_0$. Therefore, they appear as the first correction
to the linear evolution in QED in all the alternative theories,
such as the Born-Infeld theories \cite{BornInfeld,Denisov}.
In QED, $\xi_t^{QED}=\frac{7}{4}\xi_l^{QED}\equiv\frac{7}{4}\xi$,
being
\begin{equation}
\xi=\frac{8 \alpha^2 \hbar^3}{45 m_e^4 c^5}\simeq 6.7\times
10^{-30}\frac{m^3}{J}.
\label{constant_xi}
\end{equation}

This quantity has dimensions of the inverse of an energy density. This means that
significant changes with respect to linear propagation can be expected when the
electromagnetic energy
density is such that $\xi\rho$ is not very small.
While such intensities may have an astrophysical or cosmological importance,
they are not achievable in the laboratory, as we shall discuss in
the next section.

Once the electromagnetic fields are expressed in terms of the four-component gauge
field $A^\mu=(A^0,{\bf A})$ as ${\bf B}=\nabla \wedge {\bf A}$ and
${\bf E}=-c \nabla A^0-\frac{\partial {\bf A}}{\partial t}$, the equations
of motion are given by the Variational Principle:

\begin{equation}
\frac{\delta \Gamma}{\delta A^\mu}=0,
\label{var_princ_gen}
\end{equation}
where $\Gamma\equiv \int {\cal L} d^4x$ is the QED effective
action. Instead of studying the resulting equations for the fields
${\bf E}$ and ${\bf B}$, that can be found in the literature\cite{mckenna,variational},
it is more convenient to consider the equations for the gauge field components $A^\mu$.
As we discussed in Ref. \cite{prl99,tommasinietal08},
such equations admit solutions in the form of linearly polarized
waves, e.g. in the $x$ direction, with $A^0=0$ and ${\bf A}=(A,0,0)$, provided
that: $i)$ the field $A$ does not depend on the variable $x$ (a {\em transversality}
condition) and $ii)$ $A(t,y,z)$ satisfies the single equation \cite{prl99,tommasinietal08}:

\begin{eqnarray}
\label{non_lin_polarized}
&0=\partial_y^2 A+\partial_z^2 A-\partial_t^2 A   + \xi_l\epsilon_0 c^2 \left\lbrace\right.&\\ \nonumber
&\left[ \right.\left.(\partial_t A)^2-3(\partial_y A)^2 -(\partial_z A)^2 \right]\partial_y^2A  + &\\ \nonumber
&\left[\right. \left. (\partial_t A)^2 - (\partial_y A)^2 - 3 (\partial_z A)^2\right] \partial_z^2  A  - &\\ \nonumber
&\left[\right.  \left. 3(\partial_t A)^2 - (\partial_y A)^2 - (\partial_z A)^2 \right] \partial_t^2 A  + &\\ \nonumber
&4\left( \partial_z A  \partial_t A \partial_z\partial_t A- \partial_z A  \partial_y A
\partial_z\partial_y A  +  \partial_y A \partial_t A \partial_y\partial_t A\right) \left. \right\rbrace ,
&\nonumber
\end{eqnarray}
where $\partial_y \equiv \frac{\partial }{\partial y}$,
$\partial_z \equiv \frac{\partial }{\partial z}$ and $\partial_t
\equiv \frac{\partial }{c \partial t}$.

Hereafter, we will restrict our discussion to this case of linearly-polarized solution.
Note that the plane-wave
solutions of the linear Maxwell equations,
such as ${\cal A}\cos\left({\bf k}\cdot{\bf r}- \omega t\right)$,
where ${\cal A}$ is a constant, ${\bf k}=(0,k_y,k_z)$ and $\omega=c \vert {\bf k}\vert$,
are still solutions of Eq. (\ref{non_lin_polarized}). However, we expect
that the non-linear terms proportional to $\xi$, due to the QED correction,
will spoil the superposition principle. As we shall see, this will
imply a change in the propagation of crossing waves.
On the other hand, no nonlinear effect appears for a parallel beam.

Of course, this nonlinearity is not a problem for the communication system:
it can easily be seen that this vacuum effect is much smaller than
that of the atmosphere. In fact, the non trivial question is the opposite: how can we
detect such a nonlinearity? In order to have a significant effect,
we need very high fields to compensate $\xi= 6.7\times
10^{-30}{m^3}/{J}$. Which is the maximum electromagnetic
energy density that can be achieved?

\section{Ultrahigh power laser beams}
Ultra intense photon sources are available thanks to the
discovery of chirped pulse amplification (CPA)\cite{strickland85} in the
late 80's and optical parametric chirped pulse amplification (OPCPA)
\cite{dubietis92} in the 90's. Very recently, the HERCULES Laser
\cite{HERCULES}
reached a peak intensity $I\sim 2\times 10^{22}{ W}{ cm}^{-2}$,
corresponding to a value $\xi\rho\sim 4 \times 10^{-12}$ for the
product giving the importance of the nonlinear QED effects.
In few years, the projected European Extreme Light Infrastructure (ELI) \cite{ELI} will
achieve $I\sim 10^{25}W cm^{-2}$,  or $\xi\rho\sim 2 \times 10^{-9}$.
Finally, for a more distant future there is the hope to achieve $I\sim 10^{28}W cm^{-2}$ \cite{mourou06},
or  $\xi\rho\sim 2\times 10^{-6}$.

These are the highest values for the intensities of the electromagnetic
field
that are available (HERCULES) or will be available in the future.
These values are enormous. The power per pulse that has been reached at
HERCULES is $\simeq 3\times10^{14}W$,
which is an order of magnitude larger than the total power used by Mankind at present!
Of course, it is the ultrashort duration of the pulse ($\Delta t\sim 3\times 10^{-14}s$)
that make it possible to reach such intensities keeping the energy per pulse
as small as $\sim10 J$. On the other hand, at
ELI \cite{ELI}
the power per pulse will be $10^{18}W$, an order of magnitude larger
then the total power that the earth receives from the sun!
(In that case, $\Delta t\sim 10^{-18}-10^{-15}s$, so that the energy per pulse
will still be reasonably small $\sim 1-10^3 J$.)

Of course, these facilities have very important technological and scientific applications.
First, it has been proved that they can be used to produce beams of
high energy charged particles after bombarding a plasma \cite{mourou06}. This effect
is highly nonlinear, since it can produce kinetic energies of the order of
the $MeV$s or even the $GeV$s (and possibly even higher)
using the $\sim eV$ photons of the laser bumps \cite{mourou06}. This will provide
cheap and reliable sources of radiation for cancer therapy and medical diagnostics,
and may also help to ignite nuclear fusion. Moreover, it can be hoped that
this surprising new form of accelerating particles can lead to a new era in
particle physics. A second class of applications uses the small duration of the pulses,
that will allow for an ultrashort time microscopy, which can be very important
e.g. in nanotechnology.

Finally, as we shall see below, these ultrahigh power lasers
will provide a very promising tool for the exploration of
the nonlinearity of the vacuum, and in particular for the search of
$\gamma\gamma$ scattering.

\section{Previous proposals and current limit}
We have seen that the QED radiative corrections imply a nonlinear wave equations
for the vector potential $A^\mu$. On the other hand, the modified Maxwell's
equations for the average
values of the electromagnetic quantum fields ${\bf E}$ and ${\bf B}$
were obtained long ago \cite{mckenna}, and they were used to show
that the QED vacuum exhibits the DC Kerr effect, implying birefringence \cite{klein}.
More recently, several works have proposed different configurations
that can actually be used for experimental tests of the nonlinear optical response of the vacuum, e.g.
using harmonic generation in an inhomogeneous magnetic field\cite{ding92}, QED
four-wave mixing\cite{4wm}, resonant interactions in microwave cavities\cite{brodin},
or laser-induced QED vacuum birefringence\cite{Alexandrov} which can be probed by x-ray pulses\cite{xray},
among others\cite{MarklundShulka}.
All these proposals are very interesting and all of them deserve to be explored.
They will provide different, independent tests for photon-photon scattering.
However, all these cases will need some technological advance.
For instance, x-ray probing of QED birefringence requires an extra free electron laser,
which is not yet available. Other techniques like four-wave
mixing processes\cite{4wm} require the crossing of at least three beams, with
the corresponding alignment problems.
As we shall see in the next section, our proposal is
considerably simpler than the previous ones, and it
only requires a ultrahigh energy pulse.

Here, we would like to mention a very recent result obtained
by the PVLAS collaboration \cite{PVLAS}
that has searched for an evidence of birefringence of the vacuum
in a {\it magnetic} field background \cite{magnetic_bire}.
Their {\it negative} result has been used to set the {\it current limit} on the
cross section for $\gamma\gamma$ scattering at optical wavelengths.
Assuming the QED effective Lagrangian, but treating $\xi$
as a free phenomenological parameter,
their results can be translated in the limit
$\xi_{\rm exp}<3.1 \times 10^{-26}{m^3}/{J}$.
We see that this constraint is $4.6\times 10^{3}$ times higher than the QED value.
Note that in Ref. \cite{PVLAS} this result was also stated as a limit
on the photon-photon cross section,
$\sigma_{\gamma\gamma}< 4.6\times 10^{-62} m^2$, for $\lambda=1064nm$,
and
$\sigma_{\gamma\gamma}< 2.7\times 10^{-60} m^2$, for $\lambda=532nm$.

\section{A numerical solution of the QED wave equation}
In Ref. \cite{tommasinietal08} we have obtained the first
numerical solution of the full nonlinear wave equation
Eq. (\ref{non_lin_polarized}) in the case of two
counter-propagating plane waves that travel along the $z$-axis,
for simplicity having the same phase at the space-time origin.
The corresponding analytical solution of the linear wave equation (that can be obtained
by setting $\xi_l=0$ in Eq. (\ref{non_lin_polarized})) would be

\begin{eqnarray}
\label{counterprop_lin}
A_{\rm lin}(t,z)& = &\frac{\cal A}{2}\left[\cos(k z- \omega t)+\cos(k z+
\omega t)\right]\\ \nonumber
& = & {\cal A}\cos(\omega t)\cos(k z) ,
\end{eqnarray}
where ${\cal A}$ is a constant amplitude, ${\bf k}=(0,0,k)$ is the wave vector,
and $\omega=c k$ is the angular frequency. It is easy to see that Eq. (\ref{counterprop_lin})
can also be considered as the analytical solution of the linear wave equation
satisfying the boundary conditions

\begin{eqnarray}
\label{boundary_con}
A(t,0)& = &{\cal A}\cos(\omega t),\\ \nonumber
A\left(-\frac{\pi}{\omega},z\right)& = &A\left(\frac{\pi}{\omega},z\right),\\ \nonumber
\partial_z A(t,0)& = &0 ,
\end{eqnarray}
where for convenience we chose as the integration interval a `small'
cuboid of time-dimension $2 \pi/\omega$ and space-dimension $2 \pi/k$.

\begin{figure}[htb]
{\centering \resizebox{6cm}{!}{\includegraphics{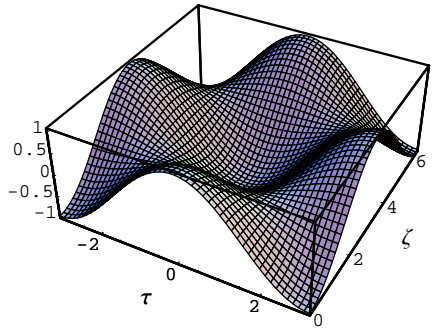}} \par}
\caption{Numerical solution $A_{\rm num}/{\cal A}$
of Eqs. (\ref{non_lin_polarized}) and (\ref{boundary_con})
for $\xi \bar\rho=0.00252$, as a function of the
adimensional time and space coordinates $\tau\equiv\omega t$ and $\zeta\equiv k z$
(from Ref. \cite{tommasinietal08}).
}
\label{fig1}
\end{figure}

\begin{figure}[htb]
{\centering \resizebox{6cm}{!}{\includegraphics{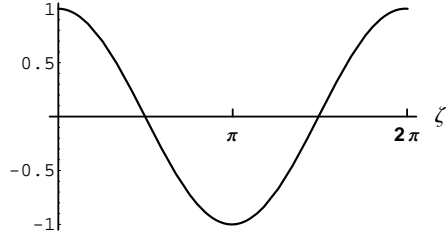}} \par}
\caption{Zero-time plot of $A_{\rm num}/{\cal A}$, as a function of the
adimensional space coordinate $\zeta\equiv k z$ (from Ref. \cite{tommasinietal08}).}
\label{fig2}
\end{figure}

\begin{figure}[htb]
{\centering \resizebox{6cm}{!}{\includegraphics{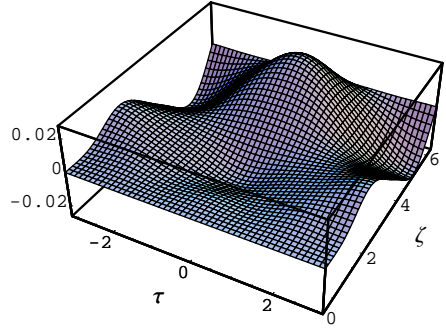}} \par}
\caption{Relative error $\left(A_{\rm num}-A_{\rm lin}\right)/{\cal A}$
of the linear approximation, as a function of the
adimensional time and space coordinates $\tau\equiv\omega t$ and $\zeta\equiv k z$
(from Ref. \cite{tommasinietal08}).}
\label{fig4}
\end{figure}

In Ref. \cite{tommasinietal08}, we have found
the solution of the nonlinear Eq. (\ref{non_lin_polarized})
that satisfies the same boundary conditions of Eq. (\ref{boundary_con}),
using the same small cuboid as the integration interval.
This solution
is shown in Figs. \ref{fig1} and \ref{fig2} for a choice of parameters
such that $\xi \epsilon_0 {\cal A}^2 \omega^2=0.01$. The corresponding
time-averaged value $\bar\rho$ of the energy density
turns out to be approximately constant
along the $z$-evolution,
giving $\xi\bar \rho=\frac{\omega}{2\pi}\xi\int_{-\pi/\omega}^{\pi/\omega}\rho dt=0.00252$.
As we have discussed above, this value of the product $\xi\rho$
is several order of magnitude larger than
what can be achieved in the laboratory in the next decades.
Of course, we will use realistic values of $\rho$ when we will
present our proposals of experiments in the last sections.
For the moment, it is interesting to note that even such
an enormous energy density is still small enough so that the
effect of the nonlinear terms gives a small
correction to the linear evolution in the short distance.
In fact, we see from Fig. \ref{fig4} that the relative difference between the linear and the
nonlinear evolution, in our integration interval which is of the order of the wavelength,
is of the order of few percent, i.e. of the same order than the adimensional
parameter $\xi \epsilon_0 {\cal A}^2 \omega^2=0.01$.
This result is not surprising, and will provide a justification for the
perturbatively-motivated variational approach that we will use in the next sections.

However, from Fig. \ref{fig4} itself, we can also appreciate that
the difference between the linear and nonlinear behavior tends to increase
along the $z$ evolution, so that it can be expected that it will eventually become large
after a distance much larger than the wavelength. In the next section, we will see an
analytical argument that confirms this expectation.

\begin{figure}[htb]
{\centering \resizebox{6cm}{!}{
\includegraphics{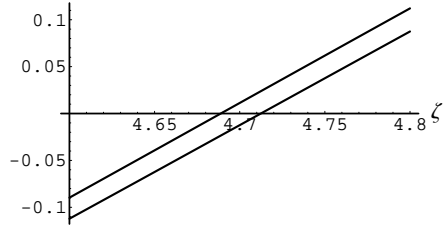}} \par}
\caption{Detail of the zero-time functions $A_{\rm num}/{\cal A}$
(upper curve) and $A_{\rm lin}/{\cal A}$
(lower curve), for values of the
adimensional space coordinate $\zeta\equiv k z$ close to the second zero
(from Ref. \cite{tommasinietal08}).}
\label{fig5}
\end{figure}

Finally, in Fig. \ref{fig5} we compare the $z$-evolution of the
solutions $A_{\rm num}$ and $A_{\rm lin}$ in a greater detail
for values of $z$ around the second zero of the solutions.
We see that $A_{\rm num}$ anticipates $A_{\rm lin}$, and the corresponding phase shift
can be evaluated numerically if we define an effective wave vector component $k_z$
by computing the value $z_0$ corresponding to $A_{\rm lin}(0,z_0)=0$, and setting $k_z z_0=3 \pi/2$.
The numerical determination of the zero gives $\zeta_0=k z_0=4.68885$, so that $k_z=1.0050 k$.
We will provide a full explanation for this result in the following section.

\section{Variational approximation}

Although it can be considered as an interesting achievement
due to its simplicity and lack of previous approximations,
in practice the numerical solution that has been discussed above
can only be obtained in the special
configuration of two counter-propagating, in-phase waves,
and for short propagation (of the order of the wavelength).
In Ref. \cite{tommasinietal08}, we have found
a variational approximation that can be used to
provide an {\it analytical}
solution which is valid even for a long evolution.
In particular, we have considered
the same two counterpropagating waves as in the previous section,
that would be described by
Eq. (\ref{counterprop_lin}) in the linear case.
Note that any of the two crossing waves, $\frac{\cal A}{2}\cos(k z- \omega t)$
and $\frac{\cal A}{2}\cos(k z+ \omega t)$, when taken alone, would be a
solution of both the linear and non-linear equations, provided that $\omega=c k$.
However, their
superposition would only solve the linear equations of motion.
Taking into account the smallness of the product $\xi\rho$,
and using the Variational Method, we have found the following analytical
approximation

\begin{equation}
A={\cal A} \cos(\omega t) \cos[(k+\chi) z],
\label{varsolution1}
\end{equation}
where $\chi=\xi_l \epsilon_0{\cal A}^2 c^2 k^3/ 2\simeq 2 \xi_l\rho k$
is a parameter that describes the
leading nonlinear effects.

As a result, we find that after a distance
$\Delta z$ the phase of the wave is shifted by a term

\begin{equation}
\Delta \phi=\chi \Delta z\simeq 2 \xi_l\rho k \Delta z
\label{dphi_headon_sym}
\end{equation}

Note also that the result of Eq. (\ref{varsolution1}) can be stated equivalently
by defining a wave vector as $k_z=k+\chi$,
that satisfies a modified dispersion relation, $\omega=c(k_z-\chi)$.

In QED ($\xi_l=\xi$), for the same choice of parameters that was considered in the previous section,
$\xi \rho\simeq\xi \epsilon_0 {\cal A}^2 \omega^2/4=0.0025$,
in agreement with the numerical simulation.
In this case, we get $\chi=0.005 k$, which is two order of magnitude smaller than $k$.
Our approximations can then be expected to be
reasonably good even for the extremely large of $\rho$ that we have
chosen here. Note also that this corresponds to a value $k_z=k+\chi=1.005 k$,
in agreement with the result that we obtained from the numerical simulation in
the previous section.

\begin{figure}[htb]
{\centering \resizebox{6cm}{!}{\includegraphics{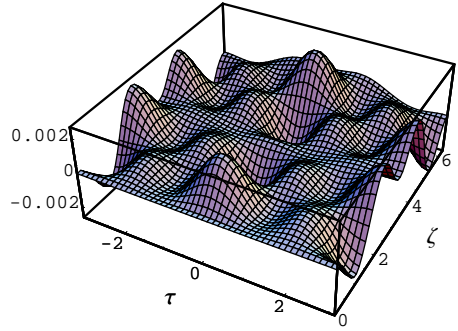}} \par}
\caption{Relative error $\left(A_{\rm num}-A_{\rm var}\right)/{\cal A}$
of the variational approximation, as a function of the
adimensional time and space coordinates $\tau\equiv\omega t$ and $\zeta\equiv k z$.}
\label{fig6}
\end{figure}

In Fig. \ref{fig6} we compare the corresponding
analytical solution $A_{\rm var}$, given in Eq. (\ref{varsolution1}),
with the numerical solution $A_{\rm num}$ of the QED wave equations
that we have found in the previous section.
Comparing with Fig. \ref{fig4},
we see that the variational solution
is an order of magnitude closer to the numerical simulation than
the linear evolution, Eq. (\ref{counterprop_lin}), that was obtained
by completely neglecting the nonlinear terms.
This is a significant improvement in such a short distance.
However, according to the previous discussion,
in the perturbative
regime corresponding to small values of the product $\xi \rho$,
the variational solution
is expected to be a good approximation even
when a longer propagation distance is considered along the $z$-axis.

On the other hand, the agreement of the variational solution with the
numerical simulation can be used as an additional, {\it a posteriori} justification
for our analytical approach.
Therefore, in Ref. \cite{tommasinietal08} we have applied the Variational Method to
different configurations, that do not allow
for a direct integration of the full Eq. (\ref{non_lin_polarized}).
The most promising one is the case of the scattering of a relatively low power
wave with a counterpropagating high power wave, both traveling along the $z$ axis.
We have also allowed for an arbitrary initial phase difference between the two waves,
assuming for simplicity that two waves have the same frequency.
In this case, the main effect will be the modification of the low power beam due to the
crossing with the high power one, which would be unaffected in a first approximation.
Using our variational approach, we have found that
the low power wave is modified, as compared to the
linear behavior, as described by the following approximated analytical solution

\begin{equation}
A_l(t,z)=\alpha_0\cos[(k+\eta) z - \omega t],
\label{A_l_sol}
\end{equation}
where $\eta=2\xi_l \epsilon_0{\cal A}^2 c^2 k^3\simeq 4 \xi\rho k$,
and $\rho$ is the (average) energy density of the high power wave.
Therefore, after a distance $\Delta z$ the low power beam accumulates
a phase shift

\begin{equation}
\Delta\Phi=\eta\Delta z\simeq 4 \xi_l\rho k \Delta z.
\label{phaseshift_lh}
\end{equation}

We stress that this result only depends on the energy density of the high power
wave, as far as it much larger than that of the low power wave. The initial phase
difference is found to be irrelevant.

\section{Proposal of experiments at HERCULES and at ELI}

In Ref. \cite{tommasinietal08}, we have proposed an
experimental setup
which can provide the simplest and most sensitive approach for
the optical search of photon-photon scattering in vacuum.
We have also showed that this porposal can be used at future exawatt facilities
such as ELI \cite{ELI} to detect the QED effect.
Here, we will briefly review that idea, and show that it
can also be used at {\it present} high power laser facilities,
such as HERCULES \cite{HERCULES}, to search for non-QED
effects, or to improve the current limit for photon-photon scattering.

The proposed experiment can be described as follows.
A common laser pulse is divided in two
beams, A and B, one of which (say A) crosses at a $180^0$ a very high power beam.
As a result, the central
part of the distribution of the beam A has acquired a phase shift $\Delta\Phi$
with respect to beam B, that has propagated freely.

In an experiment corresponding to the parameters of the ELI project in its first
step we have pulses of wavelength $\lambda=800nm$, intensity $I=10^{25}Wcm^{-2}$ and
duration $\Delta t=10fs$ which are focused in a spot of diameter $d\approx 10\mu m$.
From Eq. (\ref{phaseshift_lh}), using for $\xi_l$ the QED value of $\xi$
as given in Eq. \eqref{constant_xi}, this
results in a phase shift $\Delta\Phi\approx 2\times 10^{-7} rad$ for beam A,
which can be resolved comparing
with the beam B which was not exposed to the effects of QED vacuum. Current
techniques like spectrally resolved two-beam coupling, which can be applied for ultrashort
pulses\cite{kang97}, can be used to this purpose, being sensible
to phase shifts as small as $\Delta\Phi= 10^{-7} rad$.
It is interesting to note that the sensitivity of this method for the detection of
photon-photon scattering may be enhanced by a suitable choice of the combination
of the intensity $I$, the wavelength $\lambda$ and the time duration $\Delta t$
that enter in Eq. (\ref{phaseshift_lh}).
In fact, taking into account that $\Delta z\simeq c\Delta t$,
Eq. (\ref{phaseshift_lh}) implies that
the most sensitive experimental configuration will be that having the maximum
value of the combination $I \Delta t/\lambda$.

Moreover, in an experiment corresponding to the parameters of the HERCULES
laser \cite{HERCULES},
$\lambda=810nm$, $I=2\times10^{22}Wcm^{-2}$, $\tau=30fs$, $d\approx 0.8\mu m$,
the same configuration as above would give $\Delta\Phi\simeq10^{-9}$ (from QED),
which is possibly beyond the possibility of detection using current techniques.
However, the experiment could find evidences of non-QED $\gamma\gamma$ scattering,
or, if the result is negative, it can be used to substantially improve the current
limit. Assuming as in Ref. \cite{PVLAS} the same form of the effective
Lagrangian than in QED, and taking $\Delta\Phi= 10^{-7} rad$ as the sensitivity in
the phase shift measurement, we see that HERCULES can prove values of the
$\xi_l$ parameter down to the value $\sim5\times 10^{-28}$, which is almost two orders
of magnitude lower than PVLAS current limit. Since the scattering cross section
depends quadratically on $\xi$,
HERCULES can improve the current limit on
$\sigma_{\gamma\gamma}$ by a factor
$(3\times 10^{-26}/5\times 10^{-28})^2\simeq 4\times 10^{3}$.
This sensitivity may be increased by the upgrades of HERCULES
that are programmed for the next months. Again, the most sensitive
configuration will be that maximizing the product $I \Delta t/\lambda$.

Finally, it could be interesting to study
the possibility of further enhancing the sensitivity of our proposed
experiment by focalizing the high power pulse
in the center of a cavity, keeping the intensity on the mirrors
below the damage threshold.
In this case, the high and low power pulse might be synchronized
to cross each other several times. The number of times they meet
would then multiply the value of the phase shift.

\section{Conclusions}

We have reviewed our recent proposal
\cite{prl99,tommasinietal08} of
a very simple experiment
for the search of photon-photon scattering in vacuum.
In this proposal, a low power laser pulse
is made to cross with an ultrahigh power laser bump.
$\gamma\gamma$ scattering will then be
tested by measuring
the phase shift of the low power laser beam, e.g. by
comparing with a third low power laser beam.
Even in the first step of ELI, we have found
\cite{tommasinietal08} that the resulting
phase shift due to the QED-induced $\gamma\gamma$ scattering
will be at least
$\Delta\Phi\approx 2\times 10^{-7} rad$,
which can be measured with present technology.
Moreover, in the present work we have
also discussed the possibility of performing
our proposed experiment at present ultrahigh power laser facilities,
such as HERCULES. In particular, we have shown that HERCULES
is already able either to detect $\gamma\gamma$ scattering,
or to improve the current limit on
the cross section
$\sigma_{\gamma\gamma}$ at optical wavelengths by more than three
orders of magnitude.

{\bf Acknowledgments} D.T. thanks Padma Shukla, Lennart Stenflo, Chuan Sheng Liu, Mattias Marklund and Bengt Eliasson for their nice hospitality in Trieste.


\begin{thebibliography}{9}

\bibitem{weinberg}{S. Weinberg, {\it The Quantum Theory of Fields}, {\bf Vol. I}, Cambridge University Press (1995).}

\bibitem{tommasini0203}{D. Tommasini, J. High Energy Phys. {\bf 07 (2002)}, 039; Opt. Spectrosc. {\bf 94}, 741 (2003).}

\bibitem{BornInfeld}{M. Born, Proc. R. Soc. London {\bf 143}, 410 (1934); M. Born and L. Infeld, Proc. R. Soc. London {\bf 144}, 425 (1934).}

\bibitem{Denisov}{V. I. Denisov {\it et al.}, Phys. Rev. D {\bf 69}, 066008 (2004).}

\bibitem{phcollider}{I. Ginzburg, G. Kotkin, V. Serbo, V. Telnov, Pizma ZhETF {\bf 34}, 514 (1981); JETP Lett. {\bf 34}, 491 (1982).}

\bibitem{mourou06}{G. A. Mourou, T. Tajima, and S. V. Bulanov, Rev. Mod. Phys. {\bf 78}, 310 (2006).}

\bibitem{HERCULES}{HERCULES, V. Yanovsky et al., Optics Express {\bf 16}, 2109 (2008).}

\bibitem{ELI}{http://www.extreme-light-infraestructure.eu.}

\bibitem{Halpern}{O. Halpern, Phys. Rev. {\bf 44}, 855 (1933); H. Euler, Ann. Physik {\bf 26}, 398 (1936).}

\bibitem{Heisenberg}{W. Heisenberg and H. Euler, Z. Physik {\bf 98}, 714 (1936).}

\bibitem{mckenna}{J. McKenna and P. M. Platzman, Phys. Rev. {\bf 129}, 2354 (1963).}

\bibitem{klein}{J. J. Klein and B. P. Nigam, Phys. Rev. {\bf 135}, B1279 (1964)}

\bibitem{prl99}{A. Ferrando, H. Michinel, M. Seco, and D. Tommasini, Phys. Rev. Lett., {\bf 99}, 150404 (2007).}

\bibitem{tommasinietal08}{D. Tommasini, A. Ferrando, H. Michinel, M. Seco, Phys. Rev. A {\bf 77}, 042101 (2008).}

\bibitem{variational}{G. Brodin, et al. Phys. Lett. A {\bf 306}, 206 (2003).}

\bibitem{strickland85}{A. D. Strickland and G. Mourou, Opt. Comm. {\bf 56}, 212 (1985); P. Maine and G. Mourou, Opt. Lett. {\bf 13}, 467 (1988).}

\bibitem{dubietis92}{A. Dubietis, G. Jonusauskas, and A. Piskarskas, Opt. Comm {\bf 88}, 437 (1992).}

\bibitem{ding92}{Y. J. Ding and A. E. Kaplan, J. Nonlinear Opt. Phys. Mater. {\bf 1}, 51 (1992).}

\bibitem{4wm}{S. L. Adler, Ann. Phys. {\bf 67}, 599 (1971);
F. Moulin and D. Bernard, Opt. Comm. {\bf 164}, 137 (1999);
E. Lundstrom et al., Phys. Rev. Lett. {\bf 96}, 083602 (2006).}

\bibitem{brodin}{G. Brodin, M. Marklund, and L. Stenflo, Phys. Rev. Lett. {\bf 87}, 171801 (2001).}

\bibitem{Alexandrov}{E. B. Aleksandrov, A. A. Anselm, and A. N. Moskalev,
Zh. Eksp. Teor. Fiz. {\bf 89}, 1181 (1985)[Sov. Phys. JETP {\bf 62}, 680 (1985)].}

\bibitem{xray}{T. Heinzl et al., Opt. Comm. {\bf 267}, 318 (2006);
A. Di Piazza et al., Phys. Rev. Lett., {\bf 97}, 083603 (2006).}

\bibitem{MarklundShulka}{
M. Marklund and P. K. Shukla, Rev. Mod. Phys. {\bf 78}, 591 (2006), and references therein.}

\bibitem{PVLAS}{M.Bregant et al., PVLAS, arXiv:0805.3036 (20 May 2008).}

\bibitem{magnetic_bire}{R.Baier and P.Breitenlohner, Nuovo Cimento {\bf 47}, 261 (1967); S.L.Adler, Ann. Phys. {\bf 67}, 559 (1971); Z. Bialynicka-Birula and I. Bialynicka-Birula, Phys. Rev. D {\bf 2} 2341 (1970); E.Iacopini and E.Zavattini, Phys. lett. B {\bf 85}, 151 (1979).}

\bibitem{kang97}{I. Kang, T. Krauss, and F. Wise, Opt. Lett. {\bf 22}, 1077 (1997).}

\end{thebibliography}
\end{document}